\documentclass[12pt]{article}
\usepackage{amssymb}
\usepackage{epsfig}

\catcode`\@=11
\def\numberbysection{\@addtoreset{equation}{section}
        \def\theequation{\thesection.\arabic{equation}}}
\def\half{\frac{1}{2}}
\def\beq{\begin{equation}}
\def\eeq{\end{equation}}
\numberbysection
\begin{document}
\begin{titlepage}
\begin{center}
\hfill DFF  1/6/02 \\
\vskip 1.in
{\Large \bf Stability of the fuzzy sphere solution from matrix model}
\vskip 0.5in
P. Valtancoli
\\[.2in]
{\em Dipartimento di Fisica, Polo Scientifico Universit\'a di Firenze \\
and INFN, Sezione di Firenze (Italy)\\
Via G. Sansone 1, 50019 Sesto Fiorentino, Italy}
\end{center}
\vskip .5in
\begin{abstract}
We consider a matrix model depending on a parameter $\lambda$
which permits the fuzzy sphere as a classical background.By
expanding the bosonic matrices around this background ones
recovers a $U(1)$ ( $U(n)$ ) noncommutative gauge theory on the
fuzzy sphere. To check classical stability of this background, we
look for new classical solutions of this model and find them for
$\lambda < 1$, that make the fuzzy sphere solution unstable for
$\lambda < \half$ and stable otherwise.
\end{abstract}
\medskip
\end{titlepage}
\pagenumbering{arabic}
\section{Introduction}

The notion of quantum space or pointless geometry has been
inspired by quantum mechanics, since the notion of a point in a
quantum phase space is meaningless because of the Heisenberg
uncertainty principle.

A noncommutative space-time is defined by replacing space-time
coordinates $x^i$ by the Hermitian generators $\hat{x}^i$ of a
noncommutative $C^{*}$-algebra of " functions on space-time".
Between the motivations of a noncommutative geometry there is
surely the hope that the use of a pointless geometry in field
theory would partially eliminate the ultraviolet divergences of
quantum field theory. In practice it would be equivalent to use a
fundamental length scale below which all phenomena are ignored.

While there has been a considerable amount of work for the quantum
field theory and string theory defined on a quantum hyperplane
\cite{1}-\cite{2}-\cite{3}, there has been little understanding
for the possibility of defining non-commutativity for curved
manifolds. As a first example in this direction, noncommutative
gauge theories on a noncommutative sphere has been derived by
expanding a matrix model around its classical solution
\cite{11}-\cite{12}. The fuzzy sphere solution is considered as a
classical background, and the fluctuations on the background from
the matrices are the fields of noncommutative gauge theory. Being
the fuzzy sphere compact \cite{18}-\cite{19}-\cite{20}, it is
possible to study it with a matrix model at finite $N$, while the
quantum plane is recovered only in the $N\rightarrow \infty$ limit
\cite{13}-\cite{14}-\cite{15}-\cite{30}-\cite{31}.

Usually matrix models are obtained by the dimensional reduction
from Yang-Mills theory; between them IIB matrix model is expected
to give the constructive definition of type IIB superstring theory
\cite{4}-\cite{5}-\cite{6}-\cite{7}-\cite{8}-\cite{9}. However the
IIB matrix model has only flat noncommutative backgrounds as
classical solutions. To describe a curved space-time we need to
add a Chern-Simons term to Yang-Mills reduced model \cite{11}

\begin{equation} S = 1/g^2 Tr [ - 1/4 [ A_i, A_j ] [ A_i, A_j ]
 + 2/3 i \rho \epsilon^{ijk} A_i A_j A_k ] \label{11} \end{equation}
where $A_i$ are three $(N+1) \times (N+1)$ hermitian matrices, or
a mass term as in the model \cite{12}

\begin{equation} S' = - 1/g^2 Tr [ 1/4 [ A_i, A_j ] [ A_i, A_j ]
 + \rho^2 A_i A_i]\label{12} \end{equation}

Generally in this paper we consider an action depending on a
parameter $\lambda$ which is an interpolation between  the two
actions (\ref{11}) and (\ref{12}) :

\begin{equation} S(\lambda) = S_0 + \lambda S_1 = - 1/g^2 Tr [
1/4 [ A_i, A_j ] [ A_i, A_j ] - 2/3 i \lambda \rho \epsilon^{ijk}
A_i A_j A_k + \rho^2 ( 1- \lambda ) A_i A_i]
\label{13}\end{equation} and discuss stability of the fuzzy sphere
solution as a function of the parameter $\lambda$.

In particular we find, with an ansatz which is exhaustive  for
$N=1$, new classical solutions of this model for $\lambda <1 $ and
compare them with the fuzzy sphere to establish the minimum of the
action $S(\lambda)$. We find that for $\lambda < \half $, the
fuzzy sphere solution is unstable and that fluctuations would let
the matrix model to decade into the new classical solutions. This
could be an obstacle to the construction of the quantum theory of
the corresponding noncommutative gauge theory. At $\lambda = \half
$ the two classes of solutions coincide, and this point is
particularly symmetric. Our solution permits to deform with
continuity the fuzzy sphere solution moving from this symmetric
point. In another point, at $\lambda = -1$, our new class of
solutions reduces again to a fuzzy sphere but with a different
radius. Instead for $\lambda
>1 $ there are no new classical solutions other than the fuzzy
sphere, at least with our ansatz.

\section{Properties of the fuzzy sphere}

The fuzzy sphere \cite{10} is a noncommutative manifold
represented by the following algebra

\beq [ \hat{x}^i , \hat{x}^j ] = i \rho \epsilon^{ijk} \hat{x}^k \
\ \ \ i,j,k = 1, 2, 3  \label{21}\eeq

$\hat{x}^i$ can be represented by $(N+1) \times (N+1)$ hermitian
matrices, which can be constructed by the generators of the
$(N+1)$-dimensional irreducible representation of $SU(2)$

\beq \hat{x}^i = \rho \hat{L}^i \label{22}\eeq

The radius of the sphere is obtained by the following condition

\beq \hat{x}^i \hat{x}^i = R^2 = \rho^2 \hat{L}^i \hat{L}^i =
\rho^2 \frac{N(N+2)}{4} \label{23}\eeq

The commutative limit is realized by

\beq R = {\rm fixed} \ \ \ \rho \rightarrow 0 \ \ \ \ ( N
\rightarrow  \infty ) \label{24}\eeq

In this limit , $\hat{x}^i$ become the normal coordinates on the
sphere $x^i$:

\begin{eqnarray} x^1 & = & R sin \theta cos \phi \nonumber \\
x^2 & = & R sin \theta sin \phi \nonumber \\ x^3 & = & R cos
\theta \label{25}
\end{eqnarray}

which produces the usual metric tensor of the sphere:

\beq ds^2 = R^2 ( d\theta^2 + \sin^2 \theta d \phi^2 ) = R^2
g_{ab} d \sigma_a d \sigma_b \label{26}\eeq

The fuzzy sphere function space is finite-dimensional, in contrast
to what happens in the commutative limit. Since the coordinates
$x^i$ are substituted by hermitian matrices, the number of
independent functions on the fuzzy sphere is $(N+1)^2$, which is
exactly the number of parameters of a $(N+1) \times (N+1)$
hermitian matrix.

This cutoff on the function space can be constructed by introducing a
cutoff parameter $N$ for the angular momentum of the spherical
harmonics. An ordinary function on the sphere can be expanded in terms
of the spherical harmonics:

\beq a( \Omega ) = \sum_{l=0}^{\infty} \ \sum_{m=-l}^{l} \ a_{lm}
\ Y_{lm} ( \Omega) \label{27}\eeq

The $Y_{lm} (\Omega) $ form a basis of the classical
infinite-dimensional function space. By introducing a cutoff $N$ on
the number $l$, the number of independent functions reduces to $
\sum_{l=0}^{N} (2l+1) = {(N+1)}^2 $. To define the non commutative analogue
 of the spherical harmonics, we appeal to their classical form:

 \beq Y_{lm} = R^{-l} \sum_a f_{a_1,a_2,...,a_l}^{(lm)} \
  x^{a_1} ... x^{a_l} \label{28}\eeq

and $f_{a_1,a_2,...,a_l}$ is a traceless and symmetric tensor. The
normalization of the spherical harmonics is fixed by

\beq \int \frac{d\Omega}{4\pi} \ Y^{*}_{l' m'} Y_{lm} = \delta_{l
l'} \delta_{m m'} \label{29}\eeq

The corresponding noncommutative spherical harmonics
$\hat{Y}_{lm}$ are $ (N+1) \times ( N+1) $ hermitian matrices ,

\beq \hat{Y}_{lm} = R^{-l} \sum_{a} \ f_{a_1,a_2,..,a_l}^{(lm)} \
\hat{x}_{a_1} .... \hat{x}_{a_l} \label{210}\eeq

defined by the same symmetric tensor $ f_{a_1,a_2,..,a_l}^{(lm)}$.
A Weyl type ordering is implicit into this definition, due to the
symmetry of the indices \cite{32}.

Normalization of the noncommutative spherical harmonics is given by

\beq \frac{1}{N+1} Tr ( \hat{Y}^{\dagger}_{l' m'} \hat{Y}_{lm} ) =
\delta_{l' l} \delta_{m' m } \label{211}\eeq

An alternative definition of this cutoff on the function space can be
given by introducing a star product on the fuzzy sphere analogous to
the Moyal star product for the plane.

A matrix on the fuzzy sphere

\beq \hat{a} = \sum^{N}_{l=0} \sum^{l}_{m=-l} \ a_{lm}
\hat{Y}_{lm} \ \ \ \  \ \ a^{*}_{lm} = a_{l -m} \ \label{212}\eeq

corresponds to an ordinary function on the commutative sphere, with a
cutoff on the angular momentum:

\beq a(\Omega) = \sum^{N}_{l=0} \sum^{l}_{m=-l} \ a_{lm} Y_{lm} (
\Omega ) \label{213}\eeq

where

\beq a(\Omega) = \frac{1}{N+1} \sum^{N}_{l=0} \sum^{l}_{m=-l} \ Tr
( \hat{Y}^{\dagger}_{lm} \hat{a} ) Y_{lm} ( \Omega )
\label{214}\eeq and the ordinary product of matrices is mapped to
the star product on the commutative sphere:

\begin{eqnarray} \hat{a} \hat{b} \ & \rightarrow & \ a * b \nonumber \\
a(\Omega) * b( \Omega ) & = & \frac{1}{N+1} \sum_{l=0}^{N}
\sum_{m=-l}^{l} \ Tr ( \hat{Y}^{\dagger}_{lm} \hat{a} \hat{b} )
Y_{lm} (\Omega) \label{215}\end{eqnarray}

The cutoff in the noncommutative spherical harmonics is consistent
since the $\hat{Y}_{lm}$ form a basis of the noncommutative
Hilbert space of maps.

Derivative operators can be constructed by the adjoint action of
$\hat{L}_i$:

\beq Ad( \hat{L}_i ) \hat{a} = \sum^{N}_{l=0} \sum^{l}_{m=-l} \
a_{lm} \ [ \hat{L}_i, \hat{Y}_{lm} ] \label{216}\eeq

In the classical limit $Ad( \hat{L}_i )$ tends to the Lie derivative
on the sphere:

\beq Ad ( \hat{L}_i )  \rightarrow L_i = \frac{1}{i}
\epsilon_{ijk} x_j \partial_k \label{217}\eeq where we can expand
the classical Lie derivative $L_i$ in terms of the Killing vectors
of the sphere

\beq L_i = - i K_i^a \partial_a \label{218}\eeq

In terms of $K_i^a$ we can form the metric tensor $g_{ab} = K^i_a
K^i_b $.

In particular the analogue of the Laplacian on the fuzzy sphere is
given by :

\begin{eqnarray}
 \frac{1}{R^2} Ad ( \hat{L}^2 ) \hat{a} & = &\frac{1}{R^2} \sum^{N}_{l=0}
\sum^{l}_{m=-l} \ a_{lm} [ \hat{L}_i , [ \hat{L}_i, \hat{Y}_{lm} ]] = \nonumber \\
& = & \sum_{l=0}^{N} \sum_{m=-l}^{l} \ \frac{l(l+1)}{R^2}  a_{lm}
\hat{Y}_{lm} \label{219}\end{eqnarray}

Trace over matrices can be mapped to the integration over functions:

\beq \frac{1}{N+1} Tr ( \hat{a} ) \ \rightarrow \ \int
\frac{d\Omega}{4\pi} a( \Omega) \label{220}\eeq

\section{Gauge theory on the fuzzy sphere from matrix model}

We now recall how to recover gauge theory on the fuzzy sphere
\cite{21}-\cite{22}-\cite{23}-\cite{24}-\cite{25} by expanding
matrices around the classical solutions of the action. Consider
firstly the following action $S_0$ :

\beq S_0 = - \frac{1}{g^2} Tr ( \frac{1}{4} [ A_i, A_j ] [ A_i , A_j ]
 + \rho^2 A_i A_i ) \label{31}\eeq

 We expand the bosonic matrices $A_i$ around the classical solution (\ref{21}) as

 \beq A_i = \hat{x}_i + \rho R \hat{a}_i \label{32}\eeq

 In this way a $U(1)$ noncommutative gauge theory on the fuzzy sphere
 is introduced through the fluctuation $\hat{a}_i$. Generalizing it to $U(N)$
 gauge group is possible by changing the background classical solution:

 \beq \hat{x}_i \ \rightarrow \ \hat{x}_i \otimes 1_m \label{33}\eeq

 Therefore the fluctuations $\hat{a}_i$ are replaced as follows:

 \beq \hat{a}_i \rightarrow \sum_{a=1}^{m^2} \ \hat{a}_i^a \otimes T^a \label{34}\eeq

 where $ T^a ( a = 1,2, ...,m^2)$ denote the generators of $U(m)$.

 The action (\ref{31}) is invariant under the unitary transformation

 \beq A_i \rightarrow U^{-1} A_i U \label{35}\eeq

Since $A_i$ has the meaning of a covariant derivative as in
(\ref{32}), it is clear that gauge symmetry of the noncommutative
gauge theories is included in the unitary transformation of the
matrix model.

For an infinitesimal transformation:

\beq U \sim 1+ i \hat{\lambda} \ \ \ \ \hat{\lambda} =
\sum^N_{l=0} \sum^l_{m=-l} \ \lambda_{lm} \hat{Y}_{lm}
\label{36}\eeq

the fluctuations around the fixed background transforms as

\beq \hat{a}_1 \ \rightarrow \ \hat{a}_i - \frac{i}{R} [ \hat{L}_i
, \hat{\lambda} ] + i [ \hat{\lambda} , \hat{a}_i ] \label{37}\eeq

By using the mapping from matrices to functions, we recover the
local star-product gauge symmetry:

\beq a_i ( \Omega) \ \rightarrow \ a_i ( \Omega ) - \frac{i}{R}
L_i \lambda( \Omega ) + i [ \lambda (\Omega), a_i ( \Omega )]_{*}
\label{38}\eeq

$()_{*}$ means that the product is to be considered as a star
product.

The corresponding field strength on the sphere is given by

\begin{eqnarray} \hat{F}_{ij} & = & \frac{1}{\rho^2 R^2} ( [A_i, A_j ] - i \rho
\epsilon_{ijk} A_k ) \nonumber \\
& = & [ \frac{\hat{L}_i}{R} , \hat{a}_j ] \ - \ [
\frac{\hat{L}_j}{R} , \hat{a}_i ] + [\hat{a}_i , \hat{a}_j ] -
\frac{i}{R} \epsilon_{ijk} \hat{a}_k \label{39}\end{eqnarray}

that is mapped to the following function

\beq F_{ij} ( \Omega ) = \frac{1}{R} L_i a_j ( \Omega ) -
\frac{1}{R} L_j a_i ( \Omega ) + [ a_i (\Omega ), a_j
(\Omega)]_{*} \label{310}\eeq

$F_{ij}$ is gauge covariant even in the $U(1)$ case, as it is
manifest from the viewpoint of the matrix model.

The model contains also a scalar field which belongs to the
adjoint representation as the gauge field, and that can be defined
as:

\beq \hat{\phi} = \frac{1}{2\rho R} ( A_i A_i - \hat{x}_i
\hat{x}_i ) = \half ( \hat{x}_i \hat{a}_i + \hat{a}_i \hat{x}_i +
\rho R \hat{a}_i \hat{a}_i ) \label{311}\eeq

However, at the noncommutative level, it is impossible to
disentangle the gauge field and the scalar field which are
contained in the matrix model. Therefore only in the classical
limit the action can be interpreted as a sum of both
contributions.

\section{Commutative limit}

The action $S_0$ is mapped through the map (\ref{220}) to the
following field theory action as follows

\begin{eqnarray} S_0 & = & -
\frac{\rho^2}{4 g^2_{YM}} Tr  \int d\Omega ( F_{ij} F_{ij} )  -
\frac{3i}{2g^2_{YM}} \epsilon_{ijk} Tr \int d\Omega {( ( L_i a_j )
a_k + \frac{\rho}{3} [ a_i, a_j] a_k - \frac{i}{2} \epsilon_{ijl}
a_l a_k )}_{*} \nonumber \\ &  - & \frac{\pi }{g^2_{YM}} \frac{N
(N+2)}{2 R^2} \label{41}\end{eqnarray}

The commutative limit is realized as

\beq R = {\rm fixed} , \ \ \ g^2_{YM} = \frac{4 \pi^2 g^2}{(N+1)
\rho^4 R^2} = {\rm fixed } \ \ \ \ N \rightarrow \infty
\label{42}\eeq

In the commutative limit, the star product becomes the commutative
product.

In this limit, the scalar field $\phi$ and the gauge field are
separable from each other as in

\beq R a_i ( \Omega ) = K_i^a b_a ( \Omega ) + \frac{x_i}{R} \phi
( \Omega ) \label{43}\eeq

where $b_a$ is a gauge field on the sphere. The field strength
$F_{ij}$ can be expanded in terms of the gauge field $b_a$ and the
scalar field $\phi$ as follows :

\beq F_{ij} ( \Omega ) = \frac{1}{R^2} K_i^a K_j^b F_{ab} +
\frac{i}{R^2} \epsilon_{ijk} x_k \phi + \frac{1}{R^2} x_J K_i^a
D_a \phi - \frac{1}{R^2} x_i K_j^a D_a \phi \label{44}\eeq

where $F_{ab} = -i ( \partial_a b_b - \partial_b b_a ) + [ b_a,
b_b ]$ and $D_a = -i \partial_a + [b_a, . ]$.

The action $S_0$ is finally rewritten as:

\begin{eqnarray} S_0 & = & - \frac{1}{4 g^2_{YM} R^2} Tr \int
d\Omega ( K_i^a K_j^b K_i^c K_j^d F_{ab} F_{cd} + 2i K_i K_j^b
F_{ab} \epsilon_{ijk} \frac{x_k}{R} \phi \nonumber \\
& + & 2 K_i^a K_i^b ( D_a \phi ) ( D_b \phi ) - 2 \phi^2 ) \nonumber \\
& - & \frac{3}{2 g^2_{YM} R^2} Tr \int d\Omega ( i \epsilon_{ijk}
K_i^a K_j^b F_{ab} \frac{x_k}{R} \phi - \phi^2 )  \nonumber \\
& = & - \frac{1}{4 g^2_{YM} R^2} Tr \int d\Omega ( F_{ab} F^{ab} +
8i \frac{\epsilon^{ab}}{\sqrt{g}} F_{ab} \phi + 2 ( D_a \phi ) (
D^a \phi ) - 8 \phi^2 ) \label{45}\end{eqnarray}

Analogously the action $S_1$ is mapped to the following field
theory action as follows:

\begin{eqnarray} S_1 & = & \frac{i}{g^2_{YM}} \epsilon_{ijk} Tr \int d\Omega ( (
L_i a_j ) a_k + \frac{R}{3} [ a_i, a_j ] a_k - \frac{i}{2}
\epsilon_{ijl} a_l a_k )_{*} \nonumber  \\
& + & \frac{\pi}{3g^2_{YM}} \frac{N (N+2)}{R^2}
\label{46}\end{eqnarray}

In the commutative limit this becomes:

\begin{eqnarray} S_1
& = & \frac{1}{g^2_{YM} R^2} Tr \int d \Omega ( i \epsilon_{ijk}
K_i^a K_j^b F_{ab} \frac{x^k}{R} \phi - \phi^2 ) = \nonumber \\
& = & \frac{1}{g^2_{YM} R^2} Tr \int d\Omega \left( \frac{i
\epsilon^{ab} }{\sqrt{g}} F_{ab} \phi - \phi^2 \right) \label{47}
\end{eqnarray}

Let us notice that with $\lambda = 2$ the complete action
$S(\lambda)$ becomes

\beq S(2) = S_0 + 2 S_1 = - \frac{1}{4g^2_{YM} R^2} Tr \int
d\Omega ( F_{ab} F^{ab}  - 2 ( \partial_a \phi ) ( \partial^a \phi
)  ) \label{48}\eeq

there is no mixing term between $\phi$ and $F_{ab}$.

With $  \lambda = \frac{3}{2} $, the action is stable, with its
minimum on the fuzzy sphere:

\beq S(\frac{3}{2}) = - \frac{1}{4g^2} Tr F_{ij} F_{ij} = -
\frac{1}{4g^2} Tr ( [A_i, A_j] - i \rho \epsilon_{ijk} A_k ) (
[A_i, A_j] - i \rho \epsilon_{ijk} A_k ) \label{49}\eeq

Possible instability of the fuzzy sphere solution will be analyzed
in details in the next section, where it will be proved that it
can appear only for $\lambda <1 $.

\section{Stability of classical solutions }

Let us start with the model $S_0$. The corresponding classical
equations of motion are :

\beq [ A^j, [A^i,A^j]] + 2 \rho^2 A^i = 0 \label{51}\eeq

This type of equation of motion is considered in
\cite{16}-\cite{17}. The first immediate consequence of this
equation is that

\beq Tr ( A^i)  = 0 \label{52}\eeq

In the case of $N=1$, the general solution is then given by the
development

\beq A^i = A^i_k \frac{\sigma^k}{2} \label{53}\eeq

that can be generalized to arbitrary $N$ with the ansatz:

\beq A^i = A^i_k \hat{L}^k \label{54}\eeq

With this ansatz, the general solution is therefore given by three
vectors $A^1_i, A^2_i, A^3_i$ satisfying:

\begin{eqnarray} & A^1_i & [  2 \rho^2 - (A^2_i)^2 - (A^3_i)^2 ] + (
A^1 \cdot A^2 ) A^2_i + ( A^1 \cdot A^3 ) A^3_i = 0 \nonumber \\
& A^2_i & [  2 \rho^2 - (A^1_i)^2 - (A^3_i)^2 ] + ( A^1 \cdot A^2)
A^1_i + ( A^2 \cdot A^3 ) A^3_i = 0 \nonumber \\
& A^3_i & [  2 \rho^2 - (A^1_i)^2 - (A^2_i)^2 ] + ( A^1 \cdot A^3
) A^1_i + ( A^2 \cdot A^3 ) A^2_i = 0 \label{55}
\end{eqnarray}
It is not difficult to find a solution to this system of
equations. By multiplying the first one with $A^2_i$ and $A^3_i$
and so on we find that :

\begin{eqnarray}
( A^1 \cdot A^2 ) ( 2 \rho^2 - (A^3_i)^2 ) + ( A^1 \cdot A^3 ) (
A^2 \cdot A^3 ) & = & 0 \nonumber \\
( A^1 \cdot A^3 ) ( 2 \rho^2 - (A^2_i)^2 ) + ( A^1 \cdot A^2 ) (
A^2 \cdot A^3 ) & = & 0 \nonumber \\
( A^2 \cdot A^3 ) ( 2 \rho^2 - (A^1_i)^2 ) + ( A^1 \cdot A^2 ) (
A^1 \cdot A^3 ) & = & 0 \label{56}
\end{eqnarray}
whose solution is

\beq ( A^1 \cdot A^2 ) = ( A^1 \cdot A^3 ) = ( A^2 \cdot A^3 ) = 0
\Longrightarrow \ \ {\rm fuzzy \ sphere } \label{57}\eeq

or

\begin{eqnarray}
A^1 \cdot A^2 & = & - \sqrt{ ( 2\rho^2 - (A^1_i)^2 ) ( 2 \rho^2 -
(A^2_i)^2 ) } = - \sqrt{ ( 2\rho^2 - \alpha^2 ) ( 2 \rho^2 -
\beta^2 ) } \nonumber \\
A^1 \cdot A^3 & = & - \sqrt{ ( 2\rho^2 - (A^1_i)^2 ) ( 2 \rho^2 -
(A^3_i)^2 ) } = - \sqrt{ ( 2\rho^2 - \alpha^2 ) ( 2 \rho^2 -
\gamma^2 ) } \nonumber \\
A^2 \cdot A^3 & = & - \sqrt{ ( 2\rho^2 - (A^2_i)^2 ) ( 2 \rho^2 -
(A^3_i)^2 ) } = - \sqrt{ ( 2\rho^2 - \beta^2 ) ( 2 \rho^2 -
\gamma^2 ) } \label{58}\end{eqnarray}

Two comments are in order; firstly the argument of the square root
must be definite positive, for example :

\beq ( 2\rho^2 - \alpha^2 ) ( 2 \rho^2 - \beta^2 ) \geq 0
\label{59}\eeq

and therefore there are only two possibilities

\begin{eqnarray}
& i) & \alpha^2 \geq 2 \rho^2 \ \ \ \beta^2 \geq 2 \rho^2 \ \ \
\gamma^2 \geq 2 \rho^2 \nonumber \\
& ii) & \alpha^2 \leq 2 \rho^2 \ \ \ \beta^2 \leq 2 \rho^2 \ \ \
\gamma^2 \leq 2 \rho^2 \label{510}\end{eqnarray}

Secondly the scalar products must be less than the product of the
moduli of the vectors i.e.

\beq (A^1 \cdot A^2 )^2 \leq \alpha^2 \beta^2 \label{511}\eeq

therefore

\beq \alpha^2 + \beta^2 \geq 2 \rho^2 \ \ \ \alpha^2 + \gamma^2
\geq 2 \rho^2 \ \ \ \beta^2 + \gamma^2 \geq 2 \rho^2
\label{512}\eeq

By multiplying the first one of Eq. (\ref{55}) by $A^1_i$, the
second one by $A^2_i$ and the third one by $A^3_i$ we find :

\beq \alpha^2 + \beta^2 + \gamma^2 = 4 \rho^2 \label{513}\eeq

Therefore the possibility i) is impossible, and we are left with
possibility ii). Moreover ii) + (\ref{513}) implies (\ref{512})
automatically.

Eq. (\ref{513}) is a two-parameter solution:

\begin{eqnarray}
\alpha^2 & = & 4 \rho^2 cos^2 \theta \nonumber \\
\beta^2 & = & 4 \rho^2 sin^2 \theta cos^2 \phi \nonumber \\
\gamma^2 & = & 4 \rho^2 sin^2 \theta sin^2 \phi
\label{514}\end{eqnarray}

and ii) implies that

\beq sin^2 \theta \geq \half \ \ \  1 - \frac{1}{2 sin^2 \theta }
\leq sin^2 \phi \leq \frac{1}{2 sin^2 \theta } \label{515}\eeq

Till now we have found implicit consistency equations.

Let us now compute the system (\ref{55}) component by component.
By parameterizing:

\begin{eqnarray}
A^1 \cdot A^2 & = & \alpha \beta cos \theta_{12} \nonumber \\
A^1 \cdot A^3 & = & \alpha \gamma cos \theta_{13} \nonumber \\
A^2 \cdot A^3 & = & \beta \gamma cos \theta_{23} \label{516}
\end{eqnarray}

we can choose the first vector in a fixed direction ( $A^i$ are
invariant under the gauge transformation $U^{-1} A^i U$ )

\begin{eqnarray}
A^1_x & = & \alpha \  \ \ A^1_y = A^1_z = 0 \nonumber \\
A^2_x & = & \beta cos \theta_{12} \ \ \ A^2_y = \beta sin
\theta_{12}
\ \ \ A^2_z = 0 \nonumber \\
A^3_x & = & \gamma cos \theta_{13} \ \ \ A^3_y = \gamma sin
\theta_{13} sin \phi \ \ \ A^3_z = \gamma sin \theta_{13} cos \phi
\label{517}
\end{eqnarray}

As a consequence

\beq cos \theta_{23} = cos \theta_{12} cos \theta_{13} + sin
\theta_{12} sin \theta_{13} sin \phi \label{518}\eeq

The component by component evaluation lead us to fix completely
the solution, i.e. to find $sin \phi$ and the relative signs as
follows:

\begin{eqnarray}
cos \phi & = & 0 \ \ \ sin \phi = 1 \ \ \ \Rightarrow \theta_{23}
= \theta_{12} - \theta_{13} \nonumber \\
cos \theta_{12} & = & - \frac{1}{\alpha \beta} \sqrt{ ( 2\rho^2
-\alpha^2 ) ( 2 \rho^2 - \beta^2 ) } \ \ \ sin \theta_{12} =
\frac{1}{\alpha \beta} \sqrt{ 2\rho^2 ( \alpha^2 + \beta^2 - 2
\rho^2 )} \nonumber \\
cos \theta_{13} & = & - \frac{1}{\alpha \gamma} \sqrt{ ( 2\rho^2
-\alpha^2 ) ( 2 \rho^2 - \gamma^2 ) } \ \ \ sin \theta_{13} = -
\frac{1}{\alpha \gamma} \sqrt{ 2\rho^2 ( \alpha^2 + \gamma^2 - 2
\rho^2 )} \nonumber \\
cos \theta_{23} & = & - \frac{1}{\beta\gamma} \sqrt{ ( 2\rho^2
-\beta^2 ) ( 2 \rho^2 - \gamma^2 ) } \ \ \ sin \theta_{23} = -
\frac{1}{\beta\gamma} \sqrt{ 2\rho^2 ( \beta^2 + \gamma^2 - 2
\rho^2 )} \label{519}
\end{eqnarray}

Now we turn to the $\lambda \neq 0 $ case. The equations of motion
for the complete action $S(\lambda)$ read now

\beq [ A^j, [ A^i, A^j]] - i \rho \lambda \epsilon^{ijk} [ A^j,
A^k ] + 2 \rho^2 ( 1 - \lambda ) A^i = 0 \label{520}\eeq

Again the condition \beq Tr ( A^i ) = 0 \label{521}\eeq implies
the following ansatz

\beq A^i = A^i_k \hat{L}^k \label{522}\eeq

from which we obtain the following system

\begin{eqnarray}
( 2 \rho^2 ( 1 - \lambda ) - \beta^2 - \gamma^2 ) A^1_i + ( A^1
\cdot A^2 ) A^2_i + ( A^1 \cdot A^3 ) A^3_i + 2 \lambda \rho
\epsilon_{ijk} A^2_j A^3_k & = & 0 \nonumber \\
( 2 \rho^2 ( 1 - \lambda ) - \alpha^2 - \gamma^2 ) A^2_i + ( A^1
\cdot A^2 ) A^1_i + ( A^2 \cdot A^3 ) A^3_i + 2 \lambda \rho
\epsilon_{ijk} A^3_j A^1_k & = & 0 \nonumber \\
( 2 \rho^2 ( 1 - \lambda ) - \alpha^2 - \beta^2 ) A^3_i + ( A^1
\cdot A^3 ) A^1_i + ( A^2 \cdot A^3 ) A^2_i + 2 \lambda \rho
\epsilon_{ijk} A^1_j A^2_k & = & 0 \label{523}
\end{eqnarray}

By multiplying the first equation by $A^2_i, A^3_i$ and so on we
obtain :

\begin{eqnarray}
A^1 \cdot A^2 & = & - \sqrt{ ( 2\rho^2 ( 1 - \lambda ) - \alpha^2
) ( 2 \rho^2 ( 1-\lambda ) - \beta^2 ) } \nonumber \\
A^1 \cdot A^3 & = & - \sqrt{ ( 2\rho^2 ( 1 - \lambda ) - \alpha^2
) ( 2 \rho^2 ( 1-\lambda ) - \gamma^2 ) } \nonumber \\
A^2 \cdot A^3 & = & - \sqrt{ ( 2\rho^2 ( 1 - \lambda ) - \beta^2 )
( 2 \rho^2 ( 1-\lambda ) - \gamma^2 ) } \label{524}
\end{eqnarray}

Again the condition that the square root is positive definite
requires that :

\beq i) \alpha^2 \geq 2 \rho^2 ( 1- \lambda ) \ \ \ \beta^2 \geq 2
\rho^2 ( 1- \lambda ) \ \ \ \gamma^2 \geq 2\rho^2 ( 1 - \lambda)
\label{525} \eeq

or

\beq ii) \alpha^2 \leq 2 \rho^2 ( 1- \lambda ) \ \ \ \beta^2 \leq
2 \rho^2 ( 1- \lambda ) \ \ \ \gamma^2 \leq 2\rho^2 ( 1 -
\lambda)\label{526} \eeq

The condition that

\beq ( A^1 \cdot A^2 )^2 \leq \alpha^2 \beta^2 \ \ \ ( A^1 \cdot
A^3 )^2 \leq \alpha^2 \gamma^2 \ \ \ ( A^2 \cdot A^3 )^2 \leq
\beta^2 \gamma^2 \label{527}\eeq

implies

\beq \alpha^2 + \beta^2 \geq 2 \rho^2 ( 1 - \lambda ) \ \ \
\beta^2 + \gamma^2 \geq 2 \rho^2 ( 1- \lambda ) \ \ \ \alpha^2 +
\gamma^2 \geq 2 \rho^2 ( 1 - \lambda ) \label{528}\eeq

if and only if $\lambda \leq 1$. For $\lambda > 1$ there are no
solutions.

Let us verify the system of equations (\ref{523}) with the
parameterizations (\ref{517}). In particular the equation for
$A^2_z$ implies that:

\beq \beta \gamma cos \theta_{23} = 2 \lambda \alpha \rho sin \phi
\label{529} \eeq

Therefore

\beq sin^2 \phi = \frac{(2\rho^2 ( 1-\lambda) - \beta^2
)(2\rho^2(1-\lambda) -\gamma^2)}{(2\rho^2 ( 1-\lambda) - \beta^2
)(2\rho^2(1-\lambda) -\gamma^2)+ 4 \lambda^2 \alpha^2 \rho^2
}\label{530} \eeq

The other equations imply the following constraint ( generalizing
eq. (\ref{513}) )

\beq \alpha^2 + \beta^2 + \gamma^2 = 4 \rho^2 ( 1- \lambda) + 4
\lambda^2 \rho^2 \label{531}\eeq

and fixes all the sign as follows

\begin{eqnarray}
cos \phi & = &  - \frac{2 \lambda \alpha \rho}{\sqrt{(2\rho^2 (
1-\lambda) - \beta^2 )(2\rho^2(1-\lambda) -\gamma^2)+ 4 \lambda^2
\alpha^2 \rho^2}} \ \ \ \nonumber \\
 sin \phi & = & \sqrt{\frac{(2\rho^2 ( 1-\lambda) - \beta^2
)(2\rho^2(1-\lambda) -\gamma^2)}{(2\rho^2 ( 1-\lambda) - \beta^2
)(2\rho^2(1-\lambda) -\gamma^2)+ 4 \lambda^2 \alpha^2 \rho^2 }} \nonumber \\
cos \theta_{12} & = & - \frac{1}{\alpha \beta} \sqrt{( 2 \rho^2 (
1- \lambda) - \alpha^2 ) ( 2 \rho^2 ( 1- \lambda) - \beta^2 ) } \
\ \ \nonumber \\ sin \theta_{12} & = & \frac{1}{\alpha\beta}
\sqrt{ 2 \rho^2 ( 1-\lambda) ( \alpha^2 + \beta^2 - 2 \rho^2 (
1-\lambda))
} \nonumber \\
cos \theta_{13} & = & - \frac{1}{\alpha \gamma} \sqrt{( 2 \rho^2 (
1- \lambda) - \alpha^2 ) ( 2 \rho^2 ( 1- \lambda) - \gamma^2 ) } \
\ \ \nonumber \\ sin \theta_{13} & = & - \frac{1}{\alpha\gamma}
\sqrt{ 2 \rho^2 ( 1-\lambda) ( \alpha^2 + \gamma^2 - 2 \rho^2 (
1-\lambda)) }
\nonumber \\
cos \theta_{23} & = & - \frac{1}{\beta \gamma} \sqrt{( 2 \rho^2 (
1- \lambda) - \beta^2 ) ( 2 \rho^2 ( 1- \lambda) - \gamma^2 ) }  \nonumber \\
sin \theta_{23} & = & - \frac{1}{\beta\gamma} \sqrt{ 2 \rho^2 (
1-\lambda) ( \beta^2 + \gamma^2 - 2 \rho^2 ( 1-\lambda))
}\label{532}
\end{eqnarray}

The position $i) $ is compatible with the constraint (\ref{531})
if and only if :

\beq \alpha^2 + \beta^2 + \gamma^2 \geq 6 \rho^2 ( 1- \lambda)
\Rightarrow 2 \lambda^2 + \lambda - 1 = 2 ( \lambda + 1 ) (
\lambda - \half ) \geq 0 \label{533}\eeq

i.e. if the following condition is met

\beq \lambda < -1  \ \ {\rm or } \ \  \lambda > \half
\label{534}\eeq

The position $ii)$ is compatible if instead

\beq -1 < \lambda < \half \label{535}\eeq

The case $ \lambda = 0 $ is therefore included in the case $ii)$,
as we concluded before.

The limiting cases $\lambda = -1 $ and $\lambda = \half$ are
interesting. They corresponds to fuzzy spheres.

For $\lambda = -1$ we obtain
\begin{eqnarray}
cos \phi = 1 \ \ \  sin \theta_{12} = 1 \ \ \ sin \theta_{13} = -1
\nonumber \\
\alpha = \beta = \gamma = 2\rho \label{536}
\end{eqnarray}

and

\beq A^1 = 2 \rho L_x \ \ \ A^2 = 2 \rho L_y \ \ \ A^3 = - 2 \rho
L_z \label{537}\eeq

For $\lambda = \half$ we obtain
\begin{eqnarray}
cos \phi = -1 \ \ \ sin \theta_{12} = 1 \ \ \ sin \theta_{13} = -1
\nonumber \\
\alpha = \beta = \gamma = \rho \label{538}
\end{eqnarray}

and \beq A^i = \rho L^i \label{539}\eeq.

\section{ Computation of the action }

Let us compute the action on our new classical solutions. Firstly
we pose

\beq [ A^i , A^j ] = i \epsilon^{klm} A^i_k A^j_l \hat{L}_m
\label{61}\eeq

therefore

\begin{eqnarray} Tr [A^i, A^j] [ A^i, A^j] & = & - \epsilon^{klm}
\epsilon^{k' l' n} A^i_k A^j_l A^i_{k'} A^j_{l'} Tr ( \hat{L}^m
\hat{L}^n )
\nonumber \\
& = & - \frac{1}{3} ( ( A^i \cdot A^i )^2 - ( A^i \cdot A^j ) (
A^i \cdot A^j )) Tr ( \hat{L}_m \hat{L}_m ) \nonumber \\
& = & - \frac{8}{3} \rho^4 ( 1- \lambda ) ( 4 \lambda^2 - \lambda
+ 1 ) Tr ( \hat{L}_i \hat{L}_i) \label{62}\end{eqnarray}

The second term is

\beq - \frac{2}{3} i \lambda  \rho Tr \epsilon_{ijk} A^i A^j A^k =
\frac{2}{3} \lambda \rho Tr ( \hat{L}_i \hat{L}_i ) \epsilon^{mnp}
A^1_m A^2_n A^3_p \label{63}\eeq

It is not difficult to compute this triple vector product:

\beq \epsilon^{mnp} A^1_m A^2_n A^3_p = \alpha \beta \gamma sin
\theta_{12} sin \theta_{13} cos \phi = 4 \lambda ( 1 - \lambda
)\rho^3 \label{64}\eeq

therefore

\beq - \frac{2}{3} i \lambda \rho Tr \epsilon_{ijk} A^i A^j A^k =
\frac{8}{3} \lambda^2 ( 1- \lambda) \rho^4 Tr ( \hat{L}_i
\hat{L}_i ) \label{65}\eeq

Finally it is not difficult to evaluate the last term

\beq \rho^2 ( 1-\lambda) Tr A^i A^i = \frac{4}{3} \rho^4 ( 1-
\lambda)( \lambda^2 - \lambda + 1 ) Tr ( \hat{L}_i \hat{L}_i )
\label{66} \eeq

We have reached the following conclusion . The evaluation of this
new class of solution is independent from the two parameters
$\theta, \phi$ and depends only on $\lambda$:

\beq S(\lambda)|_{\rm new} = S_0 + \lambda S_1 = - \frac{1}{3 g^2}
\rho^4 ( 1-\lambda) ( 2 - 2 \lambda + 4 \lambda^2 ) Tr ( \hat{L}_i
\hat{L}_i ) \label{67}\eeq

Instead the action evaluated on the fuzzy sphere solution is

\beq S(\lambda)|_{\rm fuzzy \ sphere } = - \frac{\rho^4}{g^2} (
\half - \frac{\lambda}{3} ) Tr ( \hat{L}_i \hat{L}_i )
\label{68}\eeq

The condition of stability of the fuzzy sphere solution is
therefore established by this equation:

\beq S(\lambda)|_{\rm new} - S(\lambda)|_{\rm fuzzy \ sphere } =
\frac{4}{3 g^2} \rho^4 {(\lambda - \half )}^3 Tr ( \hat{L}_i
\hat{L}_i )
> 0 \label{69}\eeq

We conclude that the fuzzy sphere solution is stable for $\lambda
\geq \half $, otherwise it is unstable.

A particular mention has to be devoted to the cases $\lambda = 1/2
$ and $ \lambda = -1 $. In both cases our new classical solutions
reduce to a fuzzy sphere. However only when $\lambda = \half $
both solutions coincide , instead when $\lambda = -1 $ the fuzzy
sphere coming from our solutions has double radius and a minus
sign in the commutations relations as follows

\beq [ A_i, A_j ] = - 2i \rho \epsilon_{ijk} A_k \ \ \ \lambda =
-1 \label{610}\eeq

The point $\lambda = \half $ is particularly symmetric, as it
corresponds to the point where both solutions coincide and where
the classical instability stops.

\section{Conclusions}

In this paper we have reviewed the construction of a
noncommutative gauge theory on a fuzzy sphere starting from a
matrix model, depending on a parameter $\lambda$. We have then
studied the classical solutions of it with an ansatz, which is
exhaustive for $N=1$, and found new solutions for $\lambda < 1$,
which make unstable the fuzzy sphere background for $\lambda <
\half$. These new solutions have the nice property to be a smooth
deformation of the fuzzy sphere around the point $\lambda =
\half$, which is particular symmetric, being the confluence of the
two types of classical solutions. There is another, less symmetric
point, ( $\lambda = - 1$), where our new class of solution reduces
to a fuzzy sphere but with a different radius. It would be nice to
continue this study by analyzing the corresponding quantum theory
around the two different backgrounds, to establish if the quantum
corrections modify our classical result on stability, and by
searching for solitons solutions to the classical noncommutative
gauge theory \cite{26}-\cite{27}-\cite{28}-\cite{29}, inside the
matrix model.

\end{document}